\documentclass[5p,sort&compress]{elsarticle}

\usepackage{graphicx}
\usepackage[hidelinks]{hyperref}
\usepackage{comment}
\usepackage[version-1-compatibility]{siunitx}
\usepackage{bm,amsmath,amsfonts,braket,todonotes,multirow,dcolumn}
\usepackage{wasysym}
\usepackage[version=4]{mhchem}
\usepackage{caption}
\usepackage{subcaption}
\usepackage{cleveref}
\usepackage{float}
\usepackage[acronym]{glossaries}
\usepackage{footnote}

\DeclareSIUnit\cts{cts}

\makesavenoteenv{tabular}
\makesavenoteenv{table}

\newacronym{tem}{TEM}{transmission electron microscope}
\newacronym{stem}{STEM}{scanning-transmission electron microscope}
\newacronym{ol}{OL}{objective lens}
\newacronym{sad}{SAD}{selected area diffraction}

\newacronym{eels}{EELS}{electron energy loss spectrometry}
\newacronym{emcd}{EMCD}{electron energy-loss magnetic chiral dichroism}

\newacronym{fib}{FIB}{focused-ion-beam}

\newacronym{evb}{EVB}{electron vortex beam}
\newacronym{evbs}{EVBs}{electron vortex beams}
\newacronym{oam}{OAM}{orbital angular momentum}
\newacronym{hvm}{HVM}{holographic vortex mask}

\newacronym{fwf}{FWF}{Austrian Science Fund}

\newacronym{mc}{MC}{mode conversion}
\newacronym{hg}{HG}{Hermite-Gaussian}
\newacronym{lg}{LG}{Laguerre-Gaussian}
\newacronym{qp}{QP}{quadrupole}
\newacronym{hpp}{HPP}{Hilbert phase-plate}
\newacronym{zac}{ZAC}{metallic-glass Zirconium-Aluminium alloy}
\newacronym{adl}{ADL}{adapter lens}
\newacronym{tl1}{TL1}{transfer lens 1}
\newacronym{mcl}{MCL}{mini-condenser lens}

\begin{document}

\title{Experimental Realisation of a $\pi$/2 Vortex Mode Converter for Electrons Using a Spherical Aberration Corrector}

\author[ifp,ustem]{T.~Schachinger \corref{cor1}}
\ead{thomas.schachinger@tuwien.ac.at}
\author[ceos]{P.~Hartel}
\author[erc,rwth]{P.-H.~Lu}
\author[ustem]{S.~L\"offler}
\author[kit]{M.~Obermair}
\author[kit]{M.~Dries}
\author[kit]{D.~Gerthsen}
\author[erc]{R. E.~Dunin-Borkowski}
\author[ifp,ustem]{P.~Schattschneider}
\cortext[cor1]{Corresponding author}

\address[ifp]{Institute of Solid State Physics, TU Wien, Wiedner~Hauptstra{\ss}e~8-10, 1040 Wien, Austria}
\address[ustem]{University Service Centre for Transmission Electron Microscopy (USTEM), TU Wien, Wiedner Hauptstra{\ss}e 8-10, 1040 Wien, Austria}
\address[ceos]{CEOS Corrected Electron Optical Systems GmbH, Englerstraße 28, 69126 Heidelberg, Germany}
\address[erc]{Ernst Ruska-Centre for Microscopy and Spectroscopy with Electrons (ER-C) and Peter Grünberg Institute, Forschungszentrum Jülich, 52425 Jülich, Germany}
\address[rwth]{RWTH Aachen University, Ahornstraße 55, 52074 Aachen, Germany}
\address[kit]{Laboratorium für Elektronenmikroskopie (LEM), Karlsruher Institut für Technologie (KIT), Engesserstraße 7, 76131 Karlsruhe, Germany}

\begin{abstract}

In light optics, beams with orbital angular momentum (OAM) can be produced by employing a properly-tuned two-cylinder-lens arrangement, also called $\pi$/2 mode converter.
It is not possible to convey this concept directly to the beam in an electron microscope due to the non-existence of cylinder lenses in commercial \gls{tem}s.
A viable work-around are readily-available electron optical elements in the form of quadrupole lenses. 
In a proof-of-principle experiment in 2012, it has been shown that a single quadrupole in combination with a Hilbert phase-plate produces a spatially-confined, transient vortex mode. 

Here, an analogue to an optical $\pi$/2 mode converter is realized by repurposing a \emph{CEOS DCOR} probe corrector in an aberration corrected TEM in a way that it resembles a dual cylinder lens using two quadrupoles.
In order to verify the presence of OAM in the output beam, a fork dislocation  grating is used as an OAM analyser.
The possibility to use magnetic quadrupole fields instead of, e.g., prefabricated fork dislocation gratings to produce electron beams carrying OAM enhances the beam brightness by almost an order of magnitude and delivers switchable high-mode purity vortex beams without unwanted side-bands. 

\end{abstract}

\maketitle

\section{Introduction}

Using electron vortex beams, a multitude of new analyses and techniques have been demonstrated, including nanoparticle rotation~\cite{VerbeeckTianVanTendeloo2013,Greenberg2018}, chiral crystal structure discrimination~\cite{JuchtmansVerbeeck2015} and free electron Landau state observations~\cite{Schattschneider2014,Schachinger2015}. Potential applications include nanoscale out-of-plane magnetic measurements using interfering \gls{oam} modes~\cite{Guzzinati2019} and incident vortex-, as well as vortex filter electron energy-loss magnetic chiral dichroism setups~\cite{PohlSchneiderRuszEtAl2015,SchattschneiderLoefflerStoeger-PollachEtAl2014,SchachingerLoefflerSteiger-ThirsfeldEtAl2017}. Detecting chiral plasmon signatures in electron energy-loss spectra could lead the way to chirality discrimination of staircase nanoparticle arrangements and molecular structural fingerprinting~\cite{Asenjo-Garcia2014,HarveyPierceChessEtAl2015}.

In the last few years substantial progress has been made in the field of dynamic electron wavefront engineering in the \gls{tem}~\cite{VerbeeckBecheMuellerCasparyEtAl2018,Tavabi2020}, holding promise for quasi instantaneous and (nearly) arbitrary wavefront shaping. It opens up the way for measurement schemes where adopting the beams' geometry and phase structure to the sample characteristics and the question at hand significantly enhances the information gained from the sample~\cite{GuzzinatiBecheLourenco-MartinsEtAl2017,LourencoMartins2020}. 
Though the propositions of dynamic wavefront engineering devices using electric fields applied to nano-fabricated structures (MEMS technology) are tempting, such devices/structures suffer from limited pixel counts and fill factors (and more technically, also from pixel addressing issues)~\cite{VerbeeckBecheMuellerCasparyEtAl2018}. Thus, only partial/coarse reconstruction of the desired wavefronts is possible so far.
Optical phase plates, working either with pulsed or cw laser excitation, are definitely an interesting and fruitful new approach but nevertheless they suffer from e.g. the limited wavelength of the laser light and the need for costly and complicated external equipment\cite{Vanacore2020}.

\begin{figure}
	\centering
	\includegraphics[width=\columnwidth]{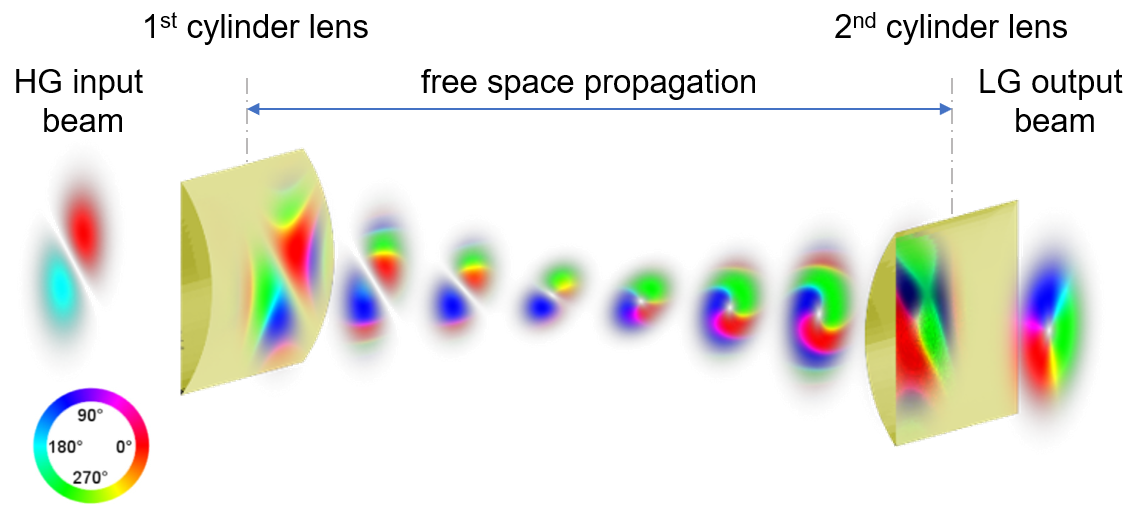} 
	\caption{Schematic diagram showing the principle of mode conversion. An incident \gls{hg}-beam is sent through a two-cylinder-lens setup at an angle of \SI{45}{\degree} with respect to the cylinder lens orientation. At this orientation, the \gls{hg}-beam resembles a superposition of two orthogonal \gls{hg} modes. The focusing action of the first cylinder lens affects only one component of the incident beam, retarding the Gouy phase evolution of the non-focused component relative to the focused one. By adjusting the input beam size to the lenses' excitations and distance, this relative Gouy phase shift can be tuned to be $\pi$/2, yielding a \gls{lg} vortex mode at the output.}
	\label{fig:MCprinciple1}
\end{figure}

Static approaches that employ prefabricated thin-film holographic devices~\cite{Bliokh2017,ShilohLuRemezEtAl2019} are practically only limited by the fabrication methods' resolution and grain size of the thin-film at hand but do not offer the possibility for dynamic reshaping of the electron beam.
A well established example of static devices are holographic fork masks for the creation of electrons carrying \gls{oam}~\cite{VerbeeckNature2010}. In the ideal case, such electron vortex beams are eigenstates of the angular momentum operator; they reveal an azimuthal phase ramp of a multiple of $2 \pi$ over the $z$-axis as depicted in Figure~\ref{fig:MCprinciple1}. 
A disadvantage of fork holograms is the low intensity and the inability to dynamically shape the beam.

An intermediate approach combining elements of both, static and dynamic, methods mentioned above has been introduced by Clark et al.~\cite{ClarkBecheGuzzinatiEtAl2013} using the combination of a static element in the form of an annular aperture and the magnetic multipole elements of a spherical aberration corrector in order to transform an incident plane wave into an electron vortex beam. Even though improved brightness and impressive resolution on the atomic scale have been demonstrated, the setup suffers from rather poor \gls{oam} mode purity.
   
Here, we present the first experimental realization of an electron optical mode conversion inside a spherical aberration corrector.
Following the pioneering idea and a preliminary description outlined by Allen et al. in 1992~\cite{AllenBeijersbergenSpreeuwEtAl1992}, mode conversion was realized in light optics by Beijersbergen et al. in 1993~\cite{Beijersbergen1993}. 
Its electron optical counterpart represents an attractive alternative to the aforementioned methods to transform an incident plane electron wave, which carries no net \gls{oam}, into an electron with $\pm \hbar$ net \gls{oam} after the mode conversion passage. 

In optics this can be readily achieved by sending a \gls{hg} beam into a set of two specifically aligned cylinder lenses. In electron optics there are, in principle, electrostatic and magnetic cylinder lenses~\cite{Rose2012}.
However, in commercial \gls{tem}s there are no such elements readily available. A viable alternative to realize cylinder lenses and, thus, mode conversion in a \gls{tem}, are magnetic quadrupole elements~\cite{Kramberger2019}, typically used as stigmators or lenses, in monochromators, energy filters and in spherical aberration correctors~\cite{Mueller2006,Krivanek2008}. Notably, a single quadrupole element has already been used to transfer vortex modes into \gls{hg} modes or vice versa~\cite{SchattschneiderStoeger-PollachVerbeeck2012,GuzzinatiClarkBecheEtAl2014,Shiloh2015}. Nonetheless, this single quadrupole approach cannot compensate for the residual astigmatic phase.       

Two magnetic quadrupoles and a static wavefunction preparation element, e.g. a \gls{hpp}\footnote{A Hilbert phase-plate is a round aperture half of which is covered by a $\pi$ phase shifting device, e.g. a thin $\alpha$-\ce{C} membrane~\cite{Danev2002}.}, can be used to build a mode conversion device that should be capable of delivering easily-switchable, singular, high-brightness and high-purity electron waves carrying \gls{oam} in multiples of $\pm \hbar$ without a residual astigmatic phase~\cite{Kramberger2019}. 

The mode conversion principle, illustrated in Figure~\ref{fig:MCprinciple1}, is based on the fact that any \gls{lg}-mode can be decomposed into a \gls{hg} mode (and vice versa) and on the Gouy phase evolution of astigmatic (electron) waves, i.e., the relative Gouy-phase difference of $\pi/2$ between the sagittal and meridional components of the electron beam upon propagation through the mode conversion device~\cite{Kramberger2019}:
\begin{equation}
\tan^{-1}\left(d/z_{rx}\right)-\tan^{-1}\left(d/z_{ry}\right)=\pi/2.
\label{eq:MC_condition_2}
\end{equation}
Here, $z_r =\frac{\pi w^2_0}{\lambda}=\frac{k w^2_0}{2}$ is the Rayleigh range of the respective electron beam component, $w_0$ represents the beam waist radius at the focus and $\lambda, k=\frac{2\pi}{\lambda}$ are the electron wavelength and wavenumber, respectively. $d$ denotes the spacing between the quadrupoles.

\begin{figure}
	\centering
	\includegraphics[width=\columnwidth]{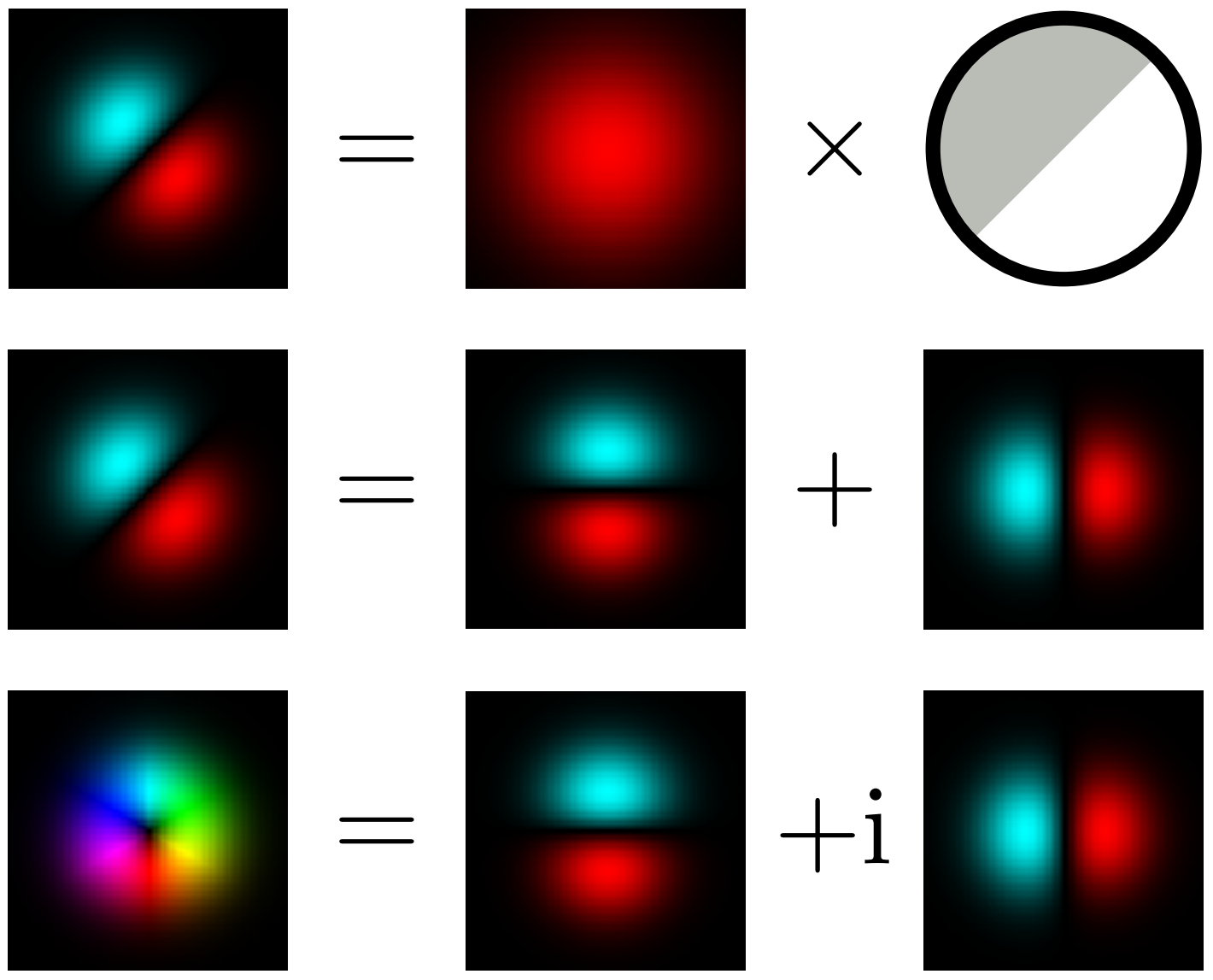} 
	\caption{Schematic mode decomposition of an \gls{lg}\textsubscript{01} mode into two \gls{hg} modes, and its preparation applying a Hilbert phase-plate to a converging Gaussian input beam far from the focus.}
	\label{fig:MCprinciple2}
\end{figure}

The desired output is a \gls{lg}\textsubscript{10} or a \gls{lg}\textsubscript{01} mode carrying an \gls{oam} of $\pm\hbar$ (using the notation of Beijersbergen~\cite{Beijersbergen1993}). 
Such a mode can be decomposed into a superposition of two \gls{hg} modes ($HG_{01}, HG_{10}$) with a relative phase shift of $\pi/2$, i.e., 
$LG_{10,01} \propto e^{-(r/w_0)^2} e^{\mp i \phi} = e^{-(r/w_0)^2}(H_1(x) \mp i H_1(y))= HG_{10}(x,y) \mp i HG_{01}(x,y)$ with $H_1(x)$ being the first Hermite polynomial and $(x,y)$ and $(r, \phi)$ being the in-plane Cartesian and polar coordinates of the beam cross section, respectively. This is also illustrated in Figure~\ref{fig:MCprinciple2} (bottom row). 
Without phase shift, the superposition of two perpendicular \gls{hg} modes resembles a \SI{45}{\degree} rotated (diagonal) \gls{hg} mode, i.e., $HG_{10,01}((x+y)/\sqrt{2},(x-y)/\sqrt{2})= e^{-(r/w_0)^2}(H_1(x) \mp H_1(y))= HG_{10}(x,y) \mp HG_{01}(x,y)$, Figure~\ref{fig:MCprinciple2} (middle row). Mode conversion between \gls{hg} and \gls{lg} modes can be achieved by changing the relative phase between the two components, according to Eq.~\ref{eq:MC_condition_2}.

Whereas in laser optics \gls{hg}\textsubscript{10} and \gls{hg}\textsubscript{01} modes can be set up routinely, electron optics does not provide this feature, at least not for the time being. A workaround to produce \gls{hg}\textsubscript{10}-type  modes in the TEM is  
 a Hilbert phase-plate intersecting a converged Gaussian beam in the far field of the cross over --- Figure~\ref{fig:MCprinciple2} (upper row). A  mode closely approximating a \gls{hg} mode rotated by \SI{45}{\degree} is produced in the cross over.

The  phase shift of $\pi/2$ of the \gls{hg}\textsubscript{10} mode is induced in the mode conversion, as a result of the two grossly different Rayleigh ranges of the two lateral beam directions, according to Equation~\ref{eq:MC_condition_2}. 
In addition to the relative phase difference of $\pi/2$, two further requirements must be met to obtain a round, stable output mode~\cite{Kramberger2019}. First, the beam waists $w_{x,y}(z)$ on the exit plane of the mode conversion must be equal:
\begin{equation}
w_x(d/2)=w_y(d/2).
\label{eq:MC_condition_1}
\end{equation}
Second, equal radii of curvature $R(z)=z(1+(z_r/z)^2)$ after the second quadrupole are required for a stigmatic output (mode matching condition):
\begin{equation}
R_{x,z}(d/2)=R_{y,z}(d/2).
\label{eq:MC_condition_3}
\end{equation}
For the symmetric $\pi/2$ mode conversion, $R_{x,z}(d/2)$ and $R_{y,z}(d/2)$ would equal $-R(-d/2)$. 
In the asymmetric case, which was chosen in the experiments, and using quadrupoles instead of cylinder lenses, the $\pi/2$ mode conversion relations are modified to~\cite{Kramberger2019}
\begin{equation}
w_i = \sqrt{\frac{2 f_i}{k}},
\label{eq:AnalyticalMC2}
\end{equation}
\begin{equation}
f_i \cdot f_o = 2\cdot d^2,
\label{eq:AnalyticalMC1}
\end{equation}
where $f_i$ and $f_o$ are the respective focal lengths of the input- and output quadrupoles and $w_i$ is the beam waist at the input of the mode conversion. With these analytical relations at hand it is straightforward to design a ray path that fulfils the mode conversion conditions and still resembles a classical \gls{stem}-like geometry, as will be shown in the following\footnote{In the special case $f_i=f_o$, i.e., a symmetric mode conversion setup, both quadrupoles would be excited to $f_{i,o} = \sqrt{2} \cdot d$.}.

\section{Experimental Setup:  The Asymmetric $\pi/2$ Mode Converter}
\label{sec:ExpSetup}

The experiments were carried out on the \emph{FEI Titan 50-300 PICO}, which is a monochromated probe-Cs-corrected and image-Cs-Cc-corrected (S)TEM instrument equipped with a high-brightness X-FEG and a GATAN OneView $4k \times 4k$ CMOS camera. The high tension was set to \SI{200}{\kilo\volt}. 

\begin{figure}
	\centering
	\includegraphics[width=\columnwidth]{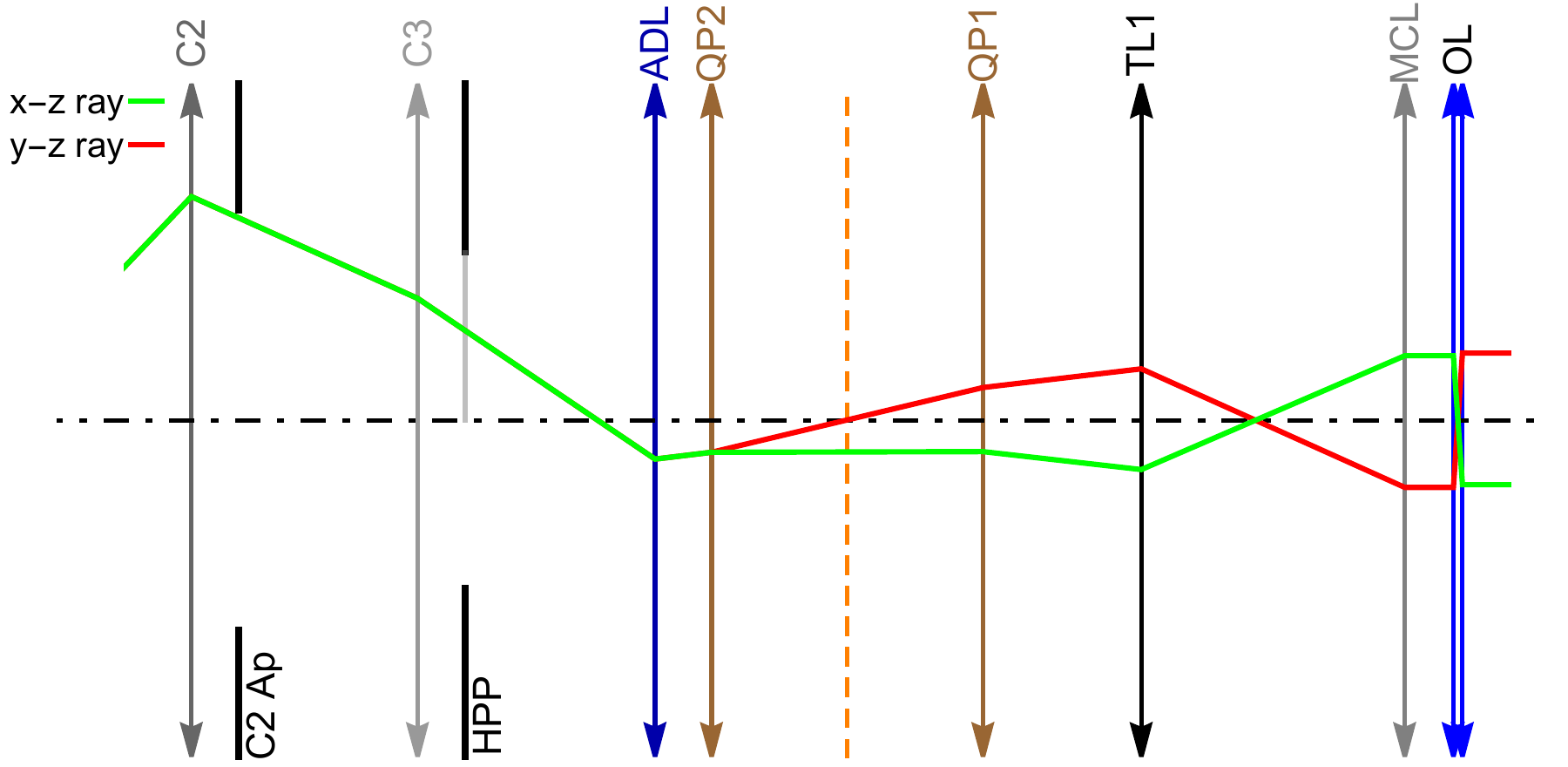} 
	\caption{Detailed ray diagram of the mode converter setup in the \emph{PICO} (S)TEM. The two hexapoles of the \emph{DCOR} are used as quadrupoles (QP1 and QP2). The \gls{adl}, in conjunction with the \emph{C2} aperture, is used to tune the input beam size and convergence angle for the mode-conversion process. The \gls{tl1} produces a cross-over in the front-focal plane of the \gls{mcl}, which is transferred to the sample plane by the mini-condensor lens and the \gls{ol}\protect\footnotemark.}
	\label{fig:MCsetup}
\end{figure}

By making use of the analytical expressions Eq.~\ref{eq:AnalyticalMC2} and Eq.~\ref{eq:AnalyticalMC1}, a ray path can be predicted that fulfils the mode conversion conditions. We used a \SI{10}{\micro\meter} \emph{C2}-aperture together with quadrupole focal lengths of $f_i=f_{QP2}=\SI{120}{\milli\meter}$, $f_o=f_{QP1}=\SI{240}{\milli\meter}$, as shown in Figure~\ref{fig:MCsetup}. 
Wave propagation simulations accounting for diffraction effects confirmed the choice of parameters~\cite{Kramberger2019}.

In passing we note that the combined action of the adapter lens (ADL) focussed on the principal plane of quadrupole QP1 and of quadrupoles QP2 and QP1 with focal lengths $f_{QP2} = f_{QP1} = \SI{120}{\mm} = d$ mimics a cylinder lens (in ray optical approximation), as evidenced in Figure~\ref{fig:MCsetup}. 
\footnotetext{This figure shows the principle geometrical ray optics beam path for $f_i=f_o=d$. Note that the actual beam path slightly deviates from the shown one. Due to the extremely low convergence angles necessary to produce Rayleigh ranges of the order of the quadrupole spacing $d$, wave optic propagation has to be applied and predicts a different quadrupole1 excitation of $f_o=2d$, see Eq.~\ref{eq:AnalyticalMC1}.}

As stated above, a further precondition for mode conversion is that the input beam must closely resemble a \gls{hg} mode, sloppily called a $\pi$-beam. 
Experimentally, the beam was produced by inserting a Hilbert phase-plate in the third condenser aperture plane (\emph{C3}) of the \emph{PICO} instrument, Figure~\ref{fig:MCsetup}. The Hilbert phase-plate was fabricated by floating-off a DC-magnetron-sputtered layer system deposited on a freshly cleaved mica substrate. The layer system consists of a \gls{zac} with \SI{11}{\nm} thickness covered by amorphous carbon (\ce{aC}) layers with \SI{6}{\nm} and \SI{12}{\nm} thickness to prevent oxidation of the \gls{zac} film. The layer system was deposited on a \gls{tem} aperture and structured to a Hilbert phase-plate by \gls{fib} milling~\cite{Dries2018}.
By tuning the film thickness, the phase of the electron wave that passes through the membrane can be shifted by $\pi$ radians for \SI{200}{\kV} electrons. 
Figures~\ref{fig:MCresults} (b) and (d) and (e) confirm that the \ce{aC/ZAC/aC} Hilbert phase-plate provides an almost symmetrical quasi-\gls{hg}-beam.

In addition to having a specific phase and intensity structure, the input $\pi$ beam must be aligned relative to the quadrupole axes. 
The precise \SI{45}{\degree} orientation of the quasi-\gls{hg}-beam relative to the quadrupole axes is essential.

\section{Results}
\label{sec:OAMAnalysis}

\begin{figure}
	\centering
	\includegraphics[width=\columnwidth]{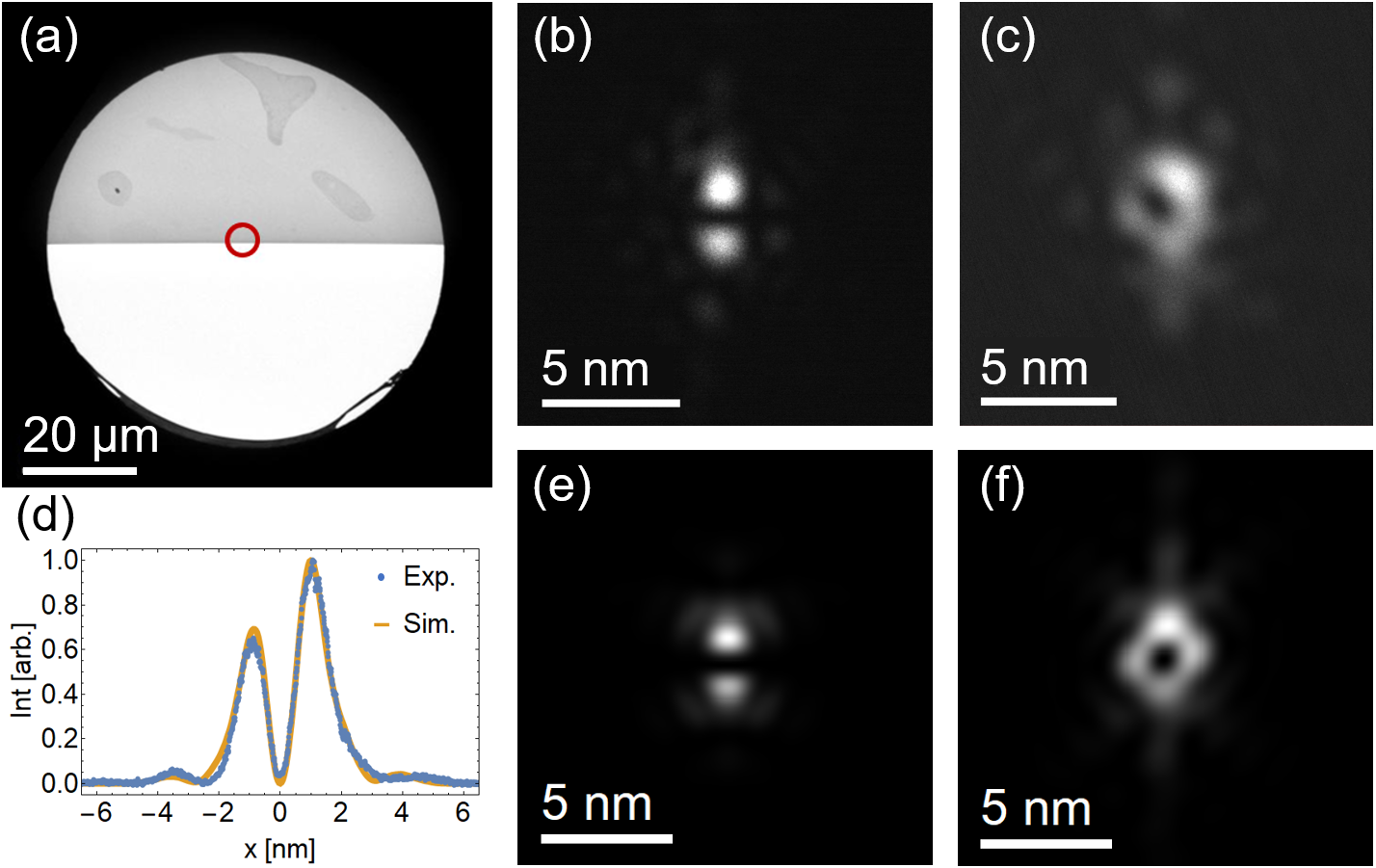} 
	\caption{(a) A layer system consisting of \ce{aC/ZAC/aC} on a \SI{70}{\micro\meter} aperture serves as a Hilbert phase-plate (HPP). The small red circle indicates the $\sim\SI{5}{\micro\meter}$ region that was illuminated during the mode conversion experiments. (b) \gls{tem} image of a focused quasi-\gls{hg}-beam produced by the Hilbert phase-plate shown in (a), demonstrating a phase shift of ~$\pi$ at an acceleration voltage of \SI{200}{\kilo\volt}, as confirmed by the simulation shown in (e) and the line profiles along the vertical direction of (b) and (e) shown in (d). (c) \gls{tem} image of the mode-converted vortex probe. (f) is the simulated vortex probe using the experimental microscope parameters.}
	\label{fig:MCresults}
\end{figure}

Figure~\ref{fig:MCresults} summarizes the experimental results (a) to (c) and compares them to simulations, (d) to (f), using the experimental parameters.
The Hilbert phase-plate, shown in Figure~\ref{fig:MCresults} (a), produces a good approximation to a $\pi$ beam (b), as it appears in the specimen plane of the objective lens with the mode conversion switched off, i.e. in the standard STEM setup. A wave optical simulation (e) and the extracted intensity profiles through the long axis of the beams (d), confirm a close to $\pi$ phase shift. The intensity difference between the two main lobes primarily stems from the slight phase shift deviation of \SI{0.11}{\pi}, from the inevitable absorption in the \gls{zac} film, which was measured to be of the order of \SI{23}{\%} and to a certain extent from alignment errors of the Hilbert phase-plate center rim. 
Note that the faint side lobes are a consequence of the sharp rim and of the irregularities in thickness and composition in the \gls{zac} film.
With the quadrupoles on, the beam changes to the ringlike structure (c) in Figure~\ref{fig:MCresults}, typical for electron waves with orbital angular momentum. In order to check where the anisotropy of the ring comes from, the output beam was simulated with a realistic, absorbing Hilbert phase-plate and a perfectly tuned mode converter. The striking agreement between simulation and experiment suggests that a) the mode conversion works as expected, and b) that the anisotropy comes from the imperfect input beam profile prepared using the Hilbert phase-plate. Side lobes seen in (b) still appear in (c), redistributed by the astigmatic transformation induced by the quadrupoles. 

\begin{figure}
	\centering
	\includegraphics[width=\columnwidth]{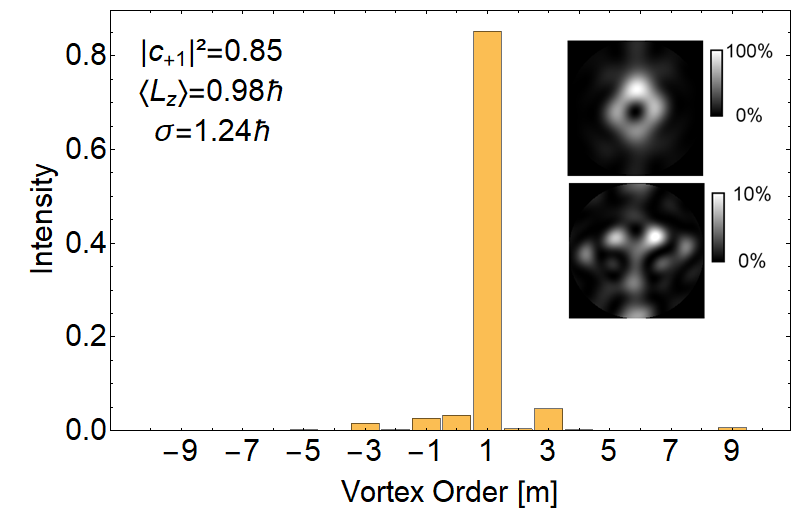} 
	\caption{\Gls{oam} spectrum of the simulated electron wave in the two quadrupole mode conversion using a Hilbert phase-plate in the condensor system of the \gls{tem}. The integration radius for calculating the \gls{oam} spectrum was chosen to contain \SI{90}{\%} of the total electron counts. The inset shows the simulated mode conversion output in the upper panel as seen at the sample plane. In the lower panel, the $m=1$ component was coherently extracted from the upper image in order to show the distribution of the $m\neq1$ components, which, in total, add \SI{\sim15}{\%} \gls{oam} impurity to the mode-converted electron probe.}
	\label{fig:OAMspectrumHPP}
\end{figure}
In order to estimate the \gls{oam} mode purity of the electron beam after mode conversion, a spectral \gls{oam} decomposition was performed, following the work of Molina–Terriza et al. and Berkhout et al.~\cite{Molina-TerrizaTorresTorner2001,BerkhoutLaveryPadgettEtAl2011}, by projecting the electron wavefunction onto vortex states $e^{i m \phi}$, where $m={..,-1,0,+1,...}$ is the topological charge or winding number. The electron wavefunction was gained by detailed image simulations using the \emph{ImageJ plugin e-beam}~\cite{Kramberger2019} and experimental microscope parameters, its absolute square is shown in Figure~\ref{fig:MCresults} (f). 
Figure~\ref{fig:OAMspectrumHPP} shows that the \gls{oam} spectrum is sharply peaked at $m=1$, indicating a successful mode conversion, despite the irregular structure of the ring. 
In practice, the useful width of vortex beams is limited by the noise level which in turn depends on the signal strength. Thus, it is instructive to choose the radius for the analysis of the \gls{oam} such that it contains \SI{90}{\%} of the total intensity. 
When choosing this radius, the mean value of the \gls{oam} is $0.98 \hbar$, the mode purity (probability of finding the $m=1$ mode) is $p_{m=1}=0.85$ and the standard deviation of the \gls{oam} spectrum is $1.24 \hbar$. By coherently extracting the $m=1$ vortex order from the simulated wave, it is possible to visualize also the weak \gls{oam} impurities, as can be seen in the inset in Figure~\ref{fig:OAMspectrumHPP}.
When taking the entire pattern the mean value of the \gls{oam} changes to $0.94 \hbar$, the mode purity is slightly reduced to $p_{m=1}=0.82$ and the standard deviation of the \gls{oam} spectrum is a bit higher $1.54 \hbar$. 
By reducing the ring radius below the \SI{90}{\%} intensity radius, the \gls{oam} spectrum becomes even sharper, due to the exclusion of features present at higher radii seen in the lower panel of Figure~\ref{fig:OAMspectrumHPP}s' inset and the standard deviation decreases. This suggests that the main source for the impurity are the higher order Fourier components caused by the sharp rim of the \gls{zac} film. This interpretation is corroborated by the fact that sending an ideal \gls{hg} beam through the mode conversion results indeed in \SI{100}{\%} mode purity.

\begin{figure}
	\centering
	\includegraphics[width=\columnwidth]{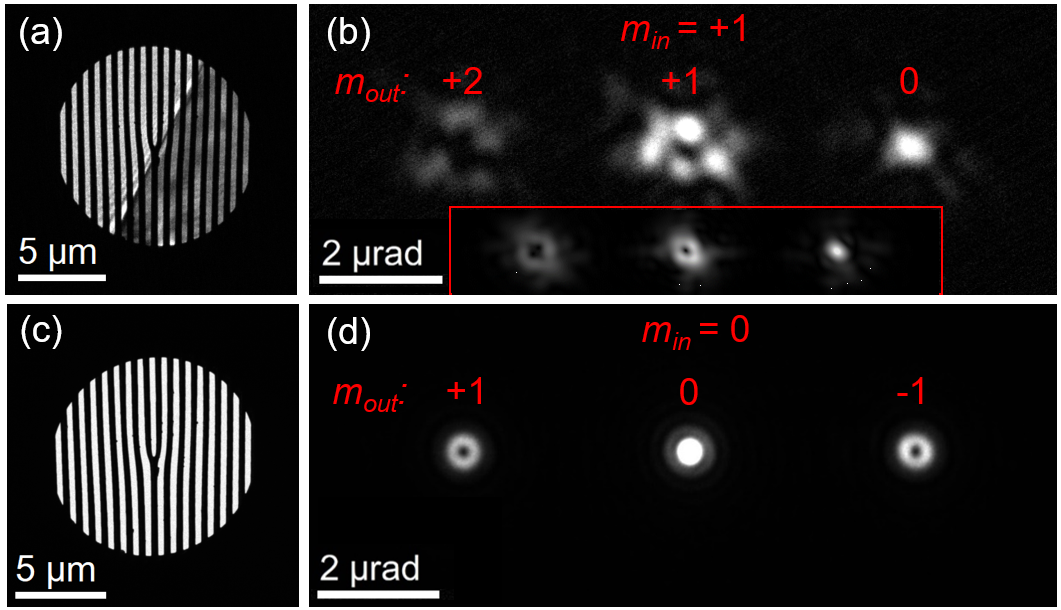} 
	\caption{Testing the \gls{oam} content of the mode-converted beam. (a) A vorticity-filtering vortex mask was placed in the sample plane. The \gls{tem} image shows the \SI{10}{\micro\meter} vortex mask together with the Hilbert phase-plate. (b) Far-field image of the mode-converted beam ($m_{in} = +1$), which was incident on the analyser mask ($m_{mask}=\{1,0,-1\}$) shown in (a). Each diffraction spot carries a topological charge of $m_{out} = m_{in} + m_{mask}$ and thus the spot pattern gets asymmetric in the case that \gls{oam} was present in the incoming electron beam. The clearly visible asymmetry in the spot pattern indicates the \gls{oam} content of the mode-converted electron beam. The inset further backs this observation by showing the result of a wave optical simulation of the \gls{oam} analysis setup using the \emph{ImageJ plugin e-beam}. For comparison we show in (c) and (d) experimental images of the action of the holographic vortex mask shown in (a) illuminated by a parallel beam without \gls{oam} content ($m_{in} = 0$) (c) resulting in a symmetric spot pattern (d).}
	\label{fig:OAMAnalyis}
\end{figure}

Additionally to the \gls{oam} analysis discussed above, a prefabricated holographic vortex mask, shown in Figure~\ref{fig:OAMAnalyis} (a) and (c), was introduced into the \gls{tem}, via the sample holder, as an \gls{oam} analyser~\cite{GuzzinatiClarkBecheEtAl2014,Shiloh2015,SchachingerLoefflerSteiger-ThirsfeldEtAl2017,SaitohHasegawaHirakawaEtAl2013,Lee2019}. Alternatively, sophisticated arrangements of \gls{oam} sorter elements, currently under development, could be employed to directly determine the \gls{oam} spectral weights~\cite{Grillo2017,Tavabi2021}.

Basically, a holographic vortex mask simply adds \gls{oam} ($m_{mask}$) to the incident electron beams' \gls{oam} ($m_{in}$) such that:  $m_{out} = m_{in} + m_{mask}$. Visually, this can be seen by an apparent asymmetry of the diffraction pattern when $m_{in}\neq0$, as each diffraction spot carries a topological charge of $m_{out}$. 
In order to detect a far-field image of the mode-converted beam going through the \gls{oam} analyser, the electron optical setup was adapted in a way to properly illuminate the holographic vortex mask and the microscope was set to the LM-diffraction mode. Figure~\ref{fig:OAMAnalyis} (b) depicts the diffraction pattern imaged in the selected area diffraction plane. As opposed to the experimental symmetric (diffraction) pattern given in Figure~\ref{fig:OAMAnalyis} (d), where $m_{in} = 0$, a central dot appears at the right hand side diffraction spot. This is a qualitative indicator for the expected \gls{oam} of the probe.

Figure~\ref{fig:OAMAnalyis} (b) reveals the clear dominance of the $m=1$ component of the mode-converted electron beam. Image simulations assuming an input beam as in Figure~\ref{fig:MCresults} (f) shown as an inset in Figure~\ref{fig:OAMAnalyis} (b) using the \emph{ImageJ plugin e-beam} confirm the presence of \gls{oam} in the probe.

\section{Discussion}
\label{sec:Discussion}

In order to put the results of the mode conversion experiments into perspective to other established electron vortex production methods Table~\ref{tab:OAMpuritycomparison} compares the essential features of vortex beams produced with different experimentally verified methods. 

The mode conversion presented here has both a high purity of \SI{\sim80}{\%} and typically only \SI{\sim10}{\%} loss of intensity because it is a single mode technique, and it provides rapid mode switching.
Reshaping the input beam into a -- or at least closer to a  \gls{hg} mode can be achieved by using e.g. binary absorptive diffractive elements, as it was done recently using soft X-rays~\cite{Lee2019}, phase masks~\cite{ShilohLereahLilachEtAl2014} or apodized apertures~\cite{Johnson2020a}.
This suppresses higher order Fourier components and would increase the purity to \SI{>95}{\%}, which is also seen in simulations. 
\begin{table*}[ht]
\caption{Comparison of the \gls{oam} mode purity of a vortex mode ($|m|=1$), the intensity loss relative to a standard round aperture, the proportion of intensity incident on the aperture redirected to the $|m|=1$ mode, the presence of singular modes and the mode switching ability of different techniques to produce electron vortex beams. For their description see the text.
1) Mode switching can be achieved, in principle, by switching the quadrupoles excitations. 2) This value is deduced from the \emph{ImageJ plugin e-beam} simulation, the value in parentheses is the measured value from~\cite{ClarkBecheGuzzinatiEtAl2014}. 3) Theoretical value taken from~\cite{HarveyPierceAgrawalEtAl2014,GrilloGazzadiKarimiEtAl2014}. 4) In principle, one can use a very small separation aperture downstream the column and retune the condenser optics to have a cross-over in the separation aperture plane in order to blank unwanted side-bands and switch between vortex orders by shifting the separation aperture or beam to the opposing vortex order~\cite{KrivanekRuszIdroboEtAl2014,Pohl2017}. 
5) Due to the lack of published mode purity values, the \gls{oam} purity and loss inside the amorphous support film was estimated using an in-house multi-slice code, see~\cite{Loeffler2019}, for \ce{Si3N4} support film thicknesses between \SI{30}{\nm} and \SI{120}{\nm}.
6) These values were calculated using reported support film thicknesses and absorption coefficient from~\cite{Johnson2020,HarveyPierceAgrawalEtAl2014,GrilloGazzadiKarimiEtAl2014}. Noteworthy, there is a strong dependence of the intensity loss on the residual supporting film thickness and imprinted topography, i.e., binary, sinusoidal or blazed gratings~\cite{Johnson2020,HarveyPierceAgrawalEtAl2014,GrilloGazzadiKarimiEtAl2014}.
7) Blazed gratings can distribute up to \SI{\sim70}{\%} to the $m=+1$ or $m=-1$ vortex mode~\cite{Johnson2020} but there are still other diffraction orders present. 
8) For binary and sinusoidal phase gratings comment number 4) applies as well, for blazed ones, this is no longer the case.  
9) The purity value is taken from~\cite{BecheWinklerPlankEtAl2016} and used for estimating the intensity redirected to the $|m|=1$ order. 10) This range is estimated using the reported thin film structure from~\cite{BecheWinklerPlankEtAl2016,ShilohLereahLilachEtAl2014} and the absorption coefficient from~\cite{Johnson2020} adjusted for material density. 
11) These values are derived from the corrector phase and simulations~\cite{ClarkBecheGuzzinatiEtAl2013}. 12) Similar to the mode conversion approach, switching of the mode can be achieved by retuning the corrector phase ramp. 
13) The purity values are deduced from the needles' phase~\cite{BecheJuchtmansVerbeeck2016}, the intensity loss is taken from~\cite{BecheJuchtmansVerbeeck2016}.
14) The switching can be done by employing a second aperture at the opposing apex of the needle or by using a special electrical feed-through holder to drive a current through a miniaturized coil in the proximity of the magnetic needle~\cite{BecheJuchtmansVerbeeck2016}. 15) \gls{oam} purity and intensity loss values have not been reported in~\cite{Tavabi2020}. 16) Fast switching and \gls{oam} modes up to $m=\pm30$ are accessible via variations of the applied voltage~\cite{Tavabi2020}.}
\label{tab:OAMpuritycomparison}
\centering
\resizebox{\textwidth}{!}{
\begin{tabular}{cccccc}
\hline
\textbf{Method} &
  \textbf{\begin{tabular}[c]{@{}c@{}}\gls{oam} purity\\  ($|m|=1$) {[}\%{]}\end{tabular}} &  \textbf{\begin{tabular}[c]{@{}c@{}}Intensity \\ loss {[}\%{]}\end{tabular}} &  \textbf{\begin{tabular}[c]{@{}c@{}}Intensity \\ $|m|=1$ {[}\%{]}\end{tabular}} & \textbf{Single mode} & \textbf{Mode switching}  \\  \hline
\multicolumn{1}{l}{Mode conversion HPP} & $\sim$85 & $\sim$11 & $\sim$76 & \ yes &  \ yes$^{1)}$ \\
\multicolumn{1}{l}{Mode conversion HG}  & $>$95 & $\sim$13 & $>$83 &  \ yes & \ yes$^{1)}$ \\
\multicolumn{1}{l}{Holographic absorption mask~\cite{ClarkBecheGuzzinatiEtAl2014,HarveyPierceAgrawalEtAl2014,GrilloGazzadiKarimiEtAl2014}} &  $>$98(70-81)$^{2)}$ & 90$^{3)}$ & 10 &  \ no$^{4)}$ & \ no$^{4)}$ \\
\multicolumn{1}{l}{Holographic phase mask~\cite{Johnson2020,HarveyPierceAgrawalEtAl2014,GrilloGazzadiKarimiEtAl2014}} &  $\sim$65-91$^{5)}$ & $\sim$62-89$^{6)}$ & $\sim$7-35 &  \ no$^{7)}$ & \ no$^{8)}$ \\
\multicolumn{1}{l}{Spiral phase mask~\cite{ShilohLereahLilachEtAl2014,BecheWinklerPlankEtAl2016}} &  $\sim$56$^{9)}$ & $\sim$48-54$^{10)}$ & $\sim$26 &  \ yes & \ no \\
\multicolumn{1}{l}{Corrector~\cite{ClarkBecheGuzzinatiEtAl2013}} &  65$^{11)}$ & 52$^{11)}$ &  31 &  \ yes  & \ yes$^{12)}$ \\
\multicolumn{1}{l}{Magnetic needle~\cite{BecheJuchtmansVerbeeck2016}} & 81-92$^{13)}$ &  1$^{13)}$ &  80-91 &  \ yes  & \ yes$^{14)}$ \\ 
\multicolumn{1}{l}{Electrostatic chopsticks~\cite{Tavabi2020}} & $-^{15)}$ &  $-^{15)}$ &  $-^{15)}$ &  \ yes  & \ yes$^{16)}$ \\ \hline
\end{tabular}
}
\end{table*}

The holographic absorption mask technique is still the gold standard for electron vortex production when it comes to mode purity. Its purity is almost \SI{100}{\%}, provided that the masks are of highest quality. But there are serious drawbacks: the reduction of the input intensity to \SI{\sim10}{\%}, the presence of other \gls{oam} modes, and the problem of switching from the $+ \hbar$  mode to the $- \hbar$ one, is only possible with field limiting apertures to blank the unwanted modes~\cite{Pohl2017}. 

Blazed holographic phase masks can distribute up to \SI{70}{\%} of the transmitted electrons to a single vortex order~\cite{Johnson2020}, which renders this approach attractive for single mode experiments, but due to elastic- and inelastic scattering in the support thin film membrane the purity- and intensity values can be significantly diminished~\cite{HarveyPierceAgrawalEtAl2014,Dries2016,Johnson2020}. Under ideal conditions blazed phase masks can deliver up to 3.5 times more electrons to a $m=\pm1$ vortex order compared to the holographic absorption masks.

Spiral phase masks distribute the incident electrons to a single mode without additional diffraction spots nearby~\cite{ShilohLereahLilachEtAl2014,BecheWinklerPlankEtAl2016}. Akin to the argument given above, electron scattering in the phase mask material poses a serious problem for the attainable mode purity and intensity, though approximately a quarter of the incident beam can be distributed to a $|m|=1$ vortex order. Mode switching can only be achieved by using separate apertures.    

Exploiting the phase tuning feature of aberration correctors a purity of \SI{65}{\%} has been reported~\cite{ClarkBecheGuzzinatiEtAl2013}. The \SI{\sim70}{\%} loss of intensity is caused by the annular aperture which selects the appropriate convergence angles needed for the optimized azimuthal phase ramp. It is a single mode technique, and mode switching is possible by retuning the corrector.

A sufficiently long and thin magnetic needle mimics a magnetic monopole, which imprints an \gls{oam} of $\pm\hbar$ on an incident plane electron wave~\cite{BecheBoxemTendelooEtAl2014,Blackburn2014,BecheJuchtmansVerbeeck2016}. A purity of \SIrange{80}{90}{\%} was reported for the magnetic needle approach. However, extreme care must be taken in order to determine the optimum magnetization and to avoid magnetic stray fields. 
Intensity loss is negligible according to a needle thickness of typically a few \SI{100}{\nm}. The single output mode can be switched by selecting the opposite needle apex, or by reversing the polarity of the needle, which needs a soft magnetic material and a miniaturized electric coil. In passing we mention that this principle is applied in squids in search of magnetic monopoles~\cite{Tassie1965}.

Recently, an electrostatic analogon to the magnetic needle approach mentioned above, using two microscopic parallel electrodes protruding into the aperture centre~\cite{Tavabi2020} was realized. 
This technique allows fast switching between $\pm \hbar$ \gls{oam} states and by varying the applied voltage, \gls{oam} modes up to $m=\pm30$ have been generated~\cite{Tavabi2020}.

\section{Conclusions and Outlook}
\label{sec:conclusion}

The \emph{DCOR} in the \emph{PICO} \gls{tem} has been reconfigured to a vortex mode converter. Approximately \SI{75}{\percent} 
of the incident electron beam was transferred to the $m=1$ vortex mode. 
This corresponds to an intensity increase by a factor of 7.5 relative to the holographic absorption mask technique and is comparable to the intensity values of the magnetic needle method. Optimizing the experimental conditions (e.g. the focal widths and the aperture defining the incident beam),  - which needs fine tuning beyond the resources available for the present experiment - the brightness and purity of the vortex beam can certainly be increased even more. 
Note that the successful production of this dual cylinder lens setup could also be used in the proposal of an anamorphotic phase plate, which is the only matter-free phase plate concept without using external laser systems~\cite{Frindt2010,Rose2010}.
Promising options currently under investigation to further improve the purity of the beam are the use of apodized masks~\cite{Johnson2020a} for a better approximation of the incident HG mode, and the replacement of the Hilbert phase-plate by a magnetic phase shifter~\cite{GuzzinatiBecheLourenco-MartinsEtAl2017}. 
An important aspect of the mode conversion technique is the ability to switch the converter from the $m=+1$ to the $m=-1$ mode. 
In theory, it is sufficient to rotate both quadrupoles by \SI{90}{\degree} in order to reverse the \gls{oam} of the output electron, which is equivalent to apply reversed currents at the quadrupoles. 
However, practice has shown that mechanical misalignment of the electron optical elements and residual aberrations stipulate more advanced retuning of the optical elements.
For applications, a user friendly interface provided by the manufacturers with pre-aligned lens and corrector currents for mode conversion is certainly feasible in the near future. 
As a consequence of using the probe correctors' quadrupoles to build the mode converter, the attainable resolution was rather poor, with the vortex diameter being \SI{2.2}{\nm}, see Figure~\ref{fig:MCresults} (c). Future efforts aim at employing quadrupole duplets upstream of the probe corrector such that the mode-converted vortex mode can be further demagnified and Cs corrected to produce an atomic sized electron vortex beam. 
Perhaps the year-long dream of routinely performing electron energy-loss magnetic chiral dichroism on single atomic columns then comes within reach. Another possible field of study deals with plasmons: Tuning the vortex size to the coherence length (usually in the nm range), rapid reversal of the \gls{oam} of a probe beam could provide new information on chirality in solid state plasmas~\cite{Asenjo-Garcia2014,HarveyPierceChessEtAl2015}.

Furthermore, a fully operational mode converter could be an incentive to implement recent proposals for quantum computation~\cite{Loeffler2019a} and quantum gate functionality~\cite{Schattschneider2020} in the electron microscope.

\section{Acknowledgements}

T.S., P.S. and S.L. acknowledge the financial support of the \gls{fwf}, projects P29687-N36 and I4309-N36.
M.D., M.O. and D.G. acknowledge funding by the German Research Foundation (Deutsche Forschungsgemeinschaft) under contract Ge 841/26.
This project has received funding from the European Research Council (ERC) under the European Union's Horizon 2020 research and innovation programme (Grant No. 856538, project “3D MAGiC”), from the European Union’s Horizon 2020 Research and Innovation Programme (Grant No. 823717, project “ESTEEM3”) and from the European Union’s Horizon 2020 Research and Innovation Programme (Grant No. 766970, project “Q-SORT”).
CEOS GmbH has received funding from the European Union’s Horizon 2020 research and innovation program under grant agreement No. 823717 – ESTEEM3. We acknowledge the support on the modification of the PICO microscope by Juri Barthel.

%\bibliography{Literature_ModeConversion_Manuscript}

\end{document}